\documentclass[10pt,twocolumn,letterpaper]{IEEEtran}

\usepackage{amsmath}
\usepackage{graphicx}
\include{psfig}
\usepackage{epsfig}
\usepackage{array}
\usepackage{psfrag}
\usepackage{amssymb}
\usepackage{color}
\usepackage{pstricks,pst-node,pst-text,pst-3d,pst-plot}
\usepackage{cite}
\usepackage{amsmath}
\usepackage{graphics}
\usepackage{trig}
\usepackage{graphicx}
\usepackage{amsmath, amsthm, amssymb}
\usepackage{anysize}
\usepackage{epsf}
\usepackage{psfrag}
\usepackage{pstricks}
\usepackage{pst-node}
\usepackage{pst-3d}
\usepackage{pst-plot}
\usepackage{epstopdf}

\usepackage{array}
\usepackage{amssymb}
\usepackage{color}

\usepackage{amsmath, amssymb, multirow, cite , epsfig, epstopdf, graphicx , graphics} 
\usepackage{graphics}
\usepackage{verbatim}
\usepackage{subfigure}
\usepackage{algorithmicx}
\usepackage{graphicx} 
\usepackage{algorithm}
\usepackage{algpseudocode}
\usepackage{algpascal}
\usepackage{algc, xcolor}
\usepackage{booktabs}

\ifCLASSINFOpdf
\else
\fi

\newcommand\AENDSKIP{\end{minipage}\bigskip}
\newcommand\AEND{\end{minipage}}

\begin{document}
\title{An Uncertainty Estimation Framework for Risk Assessment in Deep Learning-based Atrial Fibrillation Classification}


\author{
\IEEEauthorblockN{James Belen, Sajad Mousavi, Alireza Shamsoshoara, Fatemeh Afghah
\\}
\IEEEauthorblockA{School of Informatics, Computing and Cyber Systems,\\ Northern Arizona University,
Flagstaff, AZ 86011\\
}
}
\maketitle
\begin{abstract} Atrial Fibrillation (AF) is among one of the most common types of heart arrhythmia afflicting more than 3 million people in the U.S. alone. AF is estimated to be the cause of death of 1 in 4 individuals. Recent advancements in Artificial Intelligence (AI) algorithms have led to the capability of reliably detecting AF from ECG signals. While these algorithms can accurately detect AF with high precision, the discrete and deterministic classifications mean that these networks are likely to erroneously classify the given ECG signal.
This paper proposes a variational autoencoder classifier network that provides an uncertainty estimation of the network's output in addition to reliable classification accuracy. This framework can increase physicians' trust in using AI-based AF detection algorithms by providing them with a confidence score which reflects how uncertain the algorithm is about a case and recommending them to put more attention to the cases with a lower confidence score. The uncertainty is estimated by conducting multiple passes of the input through the network to build a distribution; the mean of the standard deviations is reported as the network's uncertainty. Our proposed network obtains 97.64\% accuracy in addition to reporting the uncertainty\footnote{This material is based upon work supported by the National Science Foundation under Grant Number 1657260.}.
\end{abstract}

\section{Introduction}
Atrial Fibrillation (AF) is one of the most common types of heart arrhythmia 
which can put patients at an increased risk of ischemic strokes, pulmonary strokes, and cranial strokes \cite{wolf1991atrial}. 
In 2016, close to 25,000 people had AF as the underlying cause of death with the risk of AF estimated to be about 1 in 4 individuals in the US \cite{benjamin2019heart}. Monitoring the long ECG recordings to detect the AF  cases is a time consuming task for the physicians and is subject to human errors; thus, precise and accurate automated AF detection methods can aid physicians to determine the best course of treatment for these patients. 

Artificial Neural Networks (NNs) have seen increasing use in biomedical applications \cite{yildirim2020accurate,mousavi2019inter,mousavi2019sleepeegnet,shamsoshoara2019solution, shamsoshoara2019distributed,mousavi2020single} and predicting cardiac arrhythmia such as AF.
The ``Computers in Cardiology Challenge 2001" obtained model accuracy around 80\% \cite{moody2001predicting,de2001automated,maier2001screening} with some methods achieving up to 89\% specificity and sensitivity\cite{costin2013new}. Several techniques have been developed to predict AF using deep learning approaches \cite{mousavi2020han,mousavi2019ecgnet,acharya2018automated, mousavi2020ecg}. However, these deep learning methods report their predictions with complete confidence regardless of the data. Regarding a disease with high morbidity and mortality rate as AF, having deterministic predictions from these models can cause more harm than good. A major flaw in these NNs are their likelihood at mislabeling AF as a normal rhythm as certainly as labeling a normal rhythm as AF. Thus, providing an uncertainty metric alongside the model's decision provides insight on the data and assists the physicians with reaching an accurate diagnosis \cite{wang2016towards}.

Initially, preliminary methods of uncertainty estimation involved using a softmax output probabilities as model confidence. However, the distribution of the probability using the softmax function may not accurately reflect the model's confidence as the model may report a high softmax classification, but it still retains low confidence \cite{gal2016dropout}. Several methods such as Markov Chain Monte-Carlo and variational dropout have been developed to improve this estimation the uncertainty of deep learning methods \cite{wang2016towards}. These models report a metric to assess the amount of trust that should be placed on a network's prediction. In general, there are two types of uncertainty, model or epistemic uncertainty and data or aleatoric uncertainty.  Epistemic uncertainty relates to the prediction model and its parameters. Epistemic uncertainty can be trained away as more data is given to the network, however it is much more sensitive to insufficient data. Data or aleatoric uncertainty handles the uncertainty caused by the complexity or the amount of noise present in data. Noisier or new data is likely to provide higher aleatoric uncertainties. The objective of our proposed method is to measure the aleatoric uncertainty caused by the input data data. This method minimizes the risk of misidentifying AF signals as normal and vice versa  caused by noisy data or misplaced or malfunctioning sensors. 

This paper proposes an AF prediction model using a variational autoencoder (VAE) approach to extract high level features and sample an input from a random distribution that is passed to a classifier to obtain a prediction. The use of a VAE will yield more effective uncertainty estimation as the part of the network will approximate the posterior of the input which is then sampled several times to calculate the standard deviation of this latent distribution. VAEs take advantage of the encoder-decoder structure to obtain probabilistic features like uncertainty but acts as a generative network that retains the capability of comparison to the original input. Our network uses the VAE architecture to evaluate the decision and uncertainty as each input is run through this network multiple times to obtain a distribution of decisions. The most frequent decision is taken as the model's final prediction and the standard deviation of the predictions are taken as the model's uncertainty in the decision. Because aleatoric uncertainty will reflect the amount of signal noise, ECG signals will require less pre-processing compared to other methods that use the full signal, or heart rate variability (HRV) \cite{JOO20123862}, morphological variability (MV), or wavelet features \cite{alfaras2019fast}. While deep neural networks trade off simplicity and time for better accuracy and clarity, the measure of confidence provided by more complex networks alleviate some of the trust issues present between AI and physicians. A physician may easily interpret a reasonable diagnosis with low uncertainty as correct, while a classification associated with high uncertainty may warrant a second opinion. These more complex networks provide measures to prevent misdiagnosis and clarity of decision to the physician at the cost of complexity and time.

The main contribution of this work is proposing an AF prediction model that combines state-of-the-art accuracy with uncertainty estimation that relays the confidence of our model in it's decision of the input. The model's uncertainty provides physicians a metric of trust on the predicted classification, alleviating the burden by indicating whether a case must be reviewed with expertise knowledge. 

\section{Model Description}
The proposed model is composed of three main parts, including encoder, decoder and classifier networks. The encoder network employs a residual network (ResNet) architecture to automatically extract features from the raw signal data. The residual network consists of several CNNs with shortcut layers that passes a copy of the previous output. The ResNet architecture optimizes the convolutional layers such that their outputs achieve a residual close or at zero. Having this residual retains enough of the information about the input after each CNN layer and enables the capability to build deeper networks while minimizing over-fitting issues that are common with deep CNNs.  The encoder (i.e., the ResNet network) in parallel with several dense layers encode and extract the latent variables of the data that can be used by the decoder to reconstruct the input signal.
Mathematically, the encoder approximates distribution, $q$, of input data over the latent space. The Kullback-Liebler (KL) divergence mechanism presented is the loss function used to shape the encoder to approximate the distribution (as shown in the next section). {The KL divergence terms also acts as a regularizer, preventing our network from deviating too far from approximating the posterior distribution and prevents overfitting.} Then, data points are sampled from the distribution to be fed into the decoder. 

\begin{figure*}[htbp]
    \centering
    \includegraphics[width=1.0\textwidth]{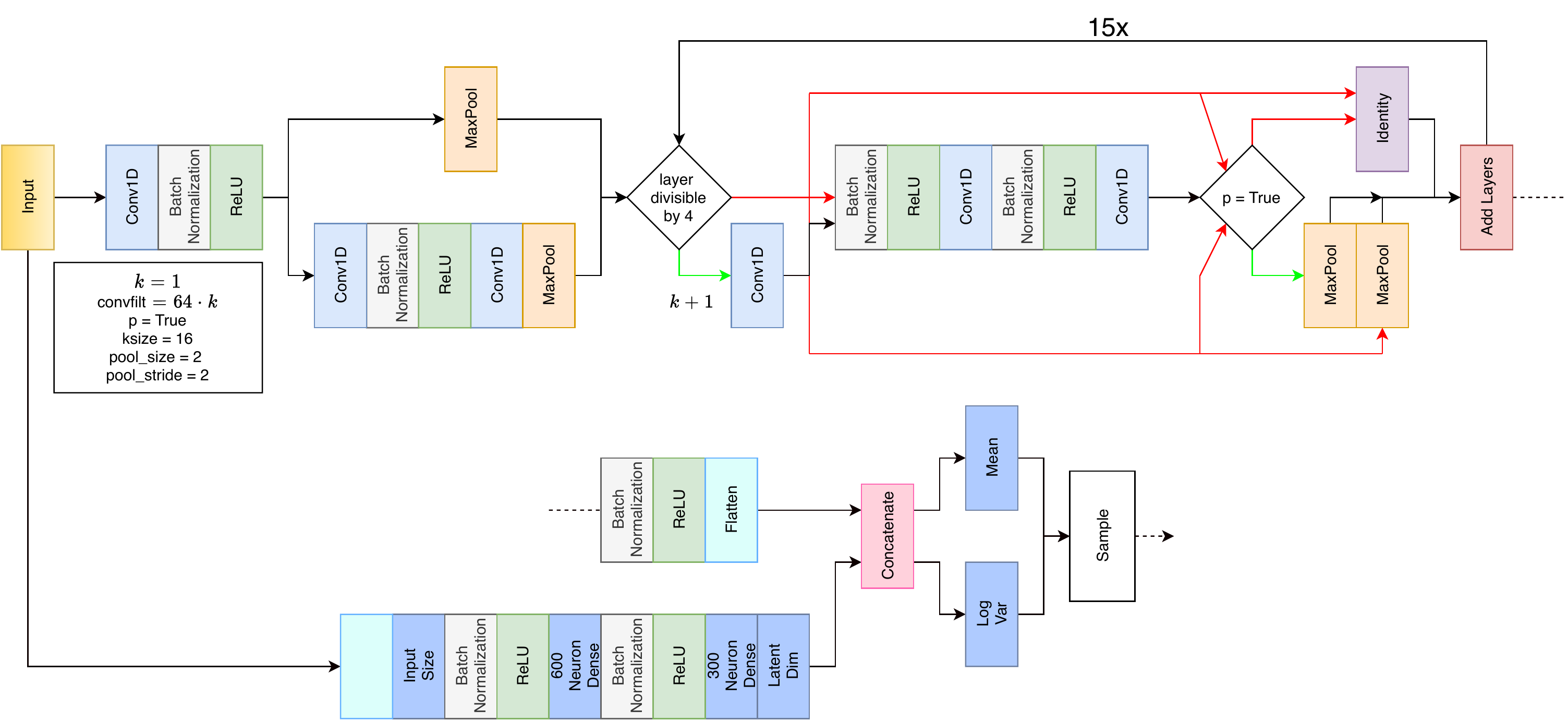}
    \centering
    \caption{Encoder Network}
    \label{fig:encoder}
\end{figure*}

\begin{figure*}[ht!]
    \centering
    \includegraphics[width=1.0\textwidth]{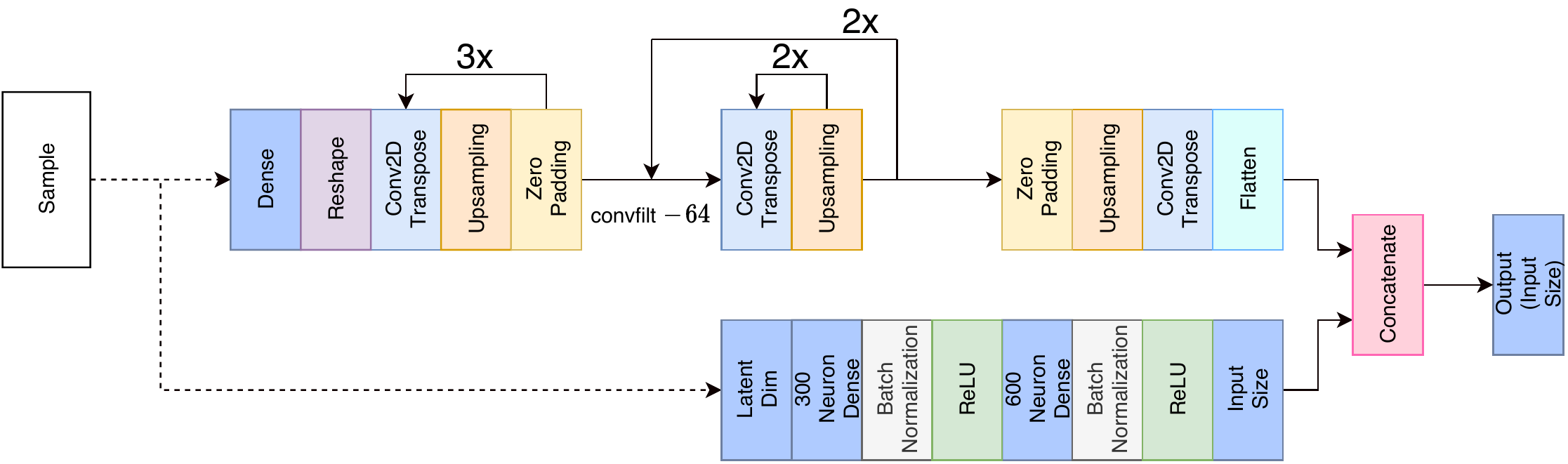}
    \centering
    \caption{Decoder Network}
    \label{fig:decoder}
\end{figure*}

\begin{figure}[ht!]
    \centering
    \includegraphics[width=0.7\linewidth]{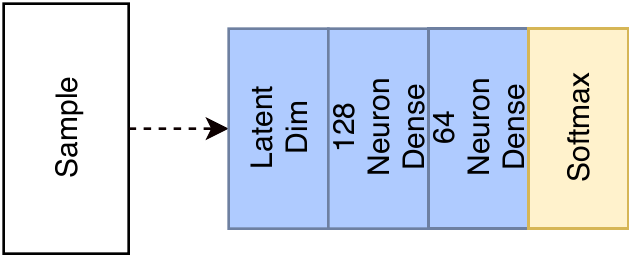}
    \centering
    \caption{Classifier Network}
    \label{fig:classifier}
\end{figure}

The decoder network consists of a network of transposed 2D convolution layers and is used to reconstruct the input data. To this end, the sampled data points are decoded by the decoder. Similar to the encoder network, the decoder network is in parallel with few dense layers and concatenated prior to a dense layer of the input size. Finally, the classifier network composing of two dense layers followed by a softmax layer employs the output of the previous step to perform the classification task. 
Figures~\ref{fig:encoder}, \ref{fig:decoder}, and \ref{fig:classifier} show more details about the encoder, decoder, and the classifier that we used in our approach accordingly.
\subsection{Uncertainty Estimation}
We obtain an uncertainty estimation of the data by implementing a variational autoencoder (VAE) network consisting of separate encoder and decoder networks. Mathematically, VAE assumes that the data is part of a posterior distribution. Using the encoder output as an approximate distribution, by training the VAE, we are able to approximate the unknown prior distribution for Bayes' theorem. The ResNet mentioned earlier acts as the encoder network for the VAE, thus we are able to infer the aleatoric uncertainty of the data by sampling features learned by the ResNet encoder as though they are random variables within a distribution. Because a layer with a random output cannot be backpropagated, discrete layers for the mean and standard deviation are used to allow backpropagation. Then the random sampling is obtained by combining the output of the random distribution output with the mean and standard deviation layers.

Because the VAE structure assumes that the input data has some posterior distribution \textit{$p(z|x)$} with \textit{$z$} being a latent variable and $x$ as the observed variable, we can obtain the posterior distribution through Bayes' Theorem.
\begin{equation}\label{eq:pzx}
    p(z|x) = \frac{p(x|z)p(z)}{p(x)}
\end{equation}
\begin{equation}\label{eq:px}
    p(x) = \int{p(x|z)p(z)dz},
\end{equation} 
however, this requires knowing or computing \textit{$p(x)$} as in \ref{eq:px}, the integral for which is intractible. On the other hand, a known distribution \textit{$q(z|x)$} can be used to approximate the posterior distribution. We use the KL Divergence to maximize the similarity. By minimizing the KL Divergence, we infer the characteristics of the posterior distribution of $x$, using our known Gaussian distribution $q$. This can be rewritten as \cite{kingma2019introduction}.
\begin{equation}\label{eq:qzx}
    E_{q(z|x)} \text{log} p(x|z) - KL(q(z|x)||p(z))
\end{equation} 
where the first term shows the reconstruction likelihood, or Evidence Lower Bound (ELBO), and the second term is the KL Divergence of $q(z|x)$ in relation to the latent distribution $p(z)$. Because the KL Divergence is always non-negative, maximizing the full equation also minimize the KL Divergence, and approximates the posterior $p(x)$. This becomes the loss function for the VAE that drives the encoder to approximate the posterior and the decoder to successfully reconstruct the input, thereby minimizing the difference between the encoder's approximate distribution and true posterior.

While learning the posterior distribution using the VAE and the latent variable $z$ poses an unsupervised learning task, adding a supervised classification output merits that features extracted in $z$ must be able to classify the inputs into one of several labels in addition to effectively reconstructing the input. To reflect this, we add a new term to the ELBO and KL Divergence such that the loss function becomes \cite{kingma2019introduction}: 
\begin{equation}\label{eq:loss}
\begin{split}
    loss = 10 * p(y|x) + E_{q(z|x)} \log p(x|z) - \\
    KL(q(z|x)||p(z)),
\end{split}
\end{equation}
where $p(y|x)$ is the error contributed by the classifier, {this term is multiplied by 10 as an unscaled value had little impact on improving the classifier network resulting in a trained encoder-decoder structure with random, untrained weights on the classifier.} This drives the VAE to extract features that can accurately reconstruct the data, and allow the added classifier network to discriminate the input into discrete classes.

After the network is trained, the output and uncertainty for the prediction is obtained by running the input through the network $n$ times. The softmax probabilities of the output are then summed and appended to an array. The classification is calculated by averaging each class's summed probability and the index of highest mean is taken as the model's classification. The uncertainty is estimated by taking the softmax probability array and calculating the variance, the square root of this variance is then taken as the standard deviation that will be used as the model's estimate of uncertainty.

\section{Experiments}
In this section, we describe the dataset along with the results of experimental analysis to evaluate the performance of this proposed method. 

\subsection{Dataset Description}
The performance of the proposed method is evaluated using the MIT-BIH AFIB database from Physionet \cite{MITBIHAt49:online}.
This dataset contains 23 patients each with two 10-hour long ECG recordings sampled at $250$ Hz over a range of $\pm 10mV$ with a $12-$bit resolution. The recordings are then split into 167,422 5-second segments and are labeled based on a threshold parameter. A segment is labeled as AF if the percentage of annotated beats exceeds the threshold parameters, anything below the threshold is considered as normal. In our method, we used a 50\% threshold, creating 66,939 AF signals and 100,483 normal signals. This caused an imbalance in the dataset, to avoid further complicating the network by creating class weights and other imbalanced data optimization, the full amount of AF signals were taken and an equal amount of normal signals were taken in order.

\subsection{Experimental Setup}
The performance of the proposed model is evaluated by comparing its classification accuracy against several state-of-the-art ANN-based AFib detection networks, and then evaluated while taking into account the uncertainty estimation.

We applied a 10-fold cross-validation training method to evaluate the proposed model. The data was split into 10 groups of data, each round involved separating the first fold for testing and training on the remaining nine.

The model was trained with a maximum of 10 epochs using a batch size of 128, with the Adam optimizer initialized at a learning rate of $a$ = 0.001. The model utilized the full ResNet model using 15 intermediate layers and the full decoder structure.The classifier loss is scaled up as, during training, the respective loss calculated by Keras did not improve the classifier's performance. Scaling this value by 10 solved the issue.

\subsection{Results} Table~\ref{tab:compare} shows the performance of our proposed model compared to several previous state-of-the-art AFIB classification methods. The table shows that our proposed network provides comparable (and in several cases better) performance in terms of the $Sensitivity$ and $Specificity$ metrics. While the overall accuracy of our method is slightly lower than the algorithm proposed by Xia et al~\cite{xia2018detecting}, its key advantage is providing a confidence metric on the network's decision. We observe that our network achieves state-of-the-art classification capability, while also providing an estimation of the aleatoric uncertainty.
\vspace{-5mm}
\begin{table}[htb]
\caption{Comparison of performance of the proposed model against other algorithms on the MIT-BIH AFIB database with the ECG segment of size 5-s.}
 \centering{
\label{tab:compare}
\resizebox{1.0\linewidth}{!}{  
\begin{tabular}{cccc}
\toprule
\textbf{} & \multicolumn{3}{c}{\textbf{Best Performance (\%)}} \\
\midrule
\textbf{Method}
&  {$Sensitivity$}  & {$Specificity$}   & {$Accuracy$}  \\
\textbf{Proposed VAE with Classifier}                                              &${97.58}$  &${97.70}$  &${97.64}$ \\
Xia et al. (2018) \cite{xia2018detecting}                         &$98.79$    &$97.87$    &$98.63$\\
Asgari et al. (2015) \cite{asgari2015automatic}                    &$97.00$    &$97.10$    &$-$\\
Lee, et al. (2013) \cite{lee2013atrial}                           &$98.20$    &$97.70$    &$-$\\
Jiang et al. (2012) \cite{jiang2012high}                          &$98.20$    &$97.50$    &$-$\\
Huang et al. (2011) \cite{huang2011novel}                         &$96.10$    &$98.10$    &$-$\\
Babaeizadeh, et al. (2009) \cite{babaeizadeh2009improvements}     &$92.00$    &$95.50$    &$-$\\
Dash et al. (2009) \cite{dash2009automatic}                        &$94.40$    &$95.10$    &$-$\\
Tateno et al. (2001) \cite{tateno2001automatic}                   &$94.40$    &$97.20$    &$-$\\
 \bottomrule  
\end{tabular} 
}
}
\end{table}
\vspace{1mm}
\section{Uncertainty Results}
Our uncertainty algorithm was tested using the same MIT-BIH AFib dataset cross-validation train and test sets. The train set is the same dataset that was used to train that iteration of the model while the test set was the same used to evaluate the model's validation metrics. 

\begin{figure}[htbp]
    \centering
    \includegraphics[width=0.488\textwidth]{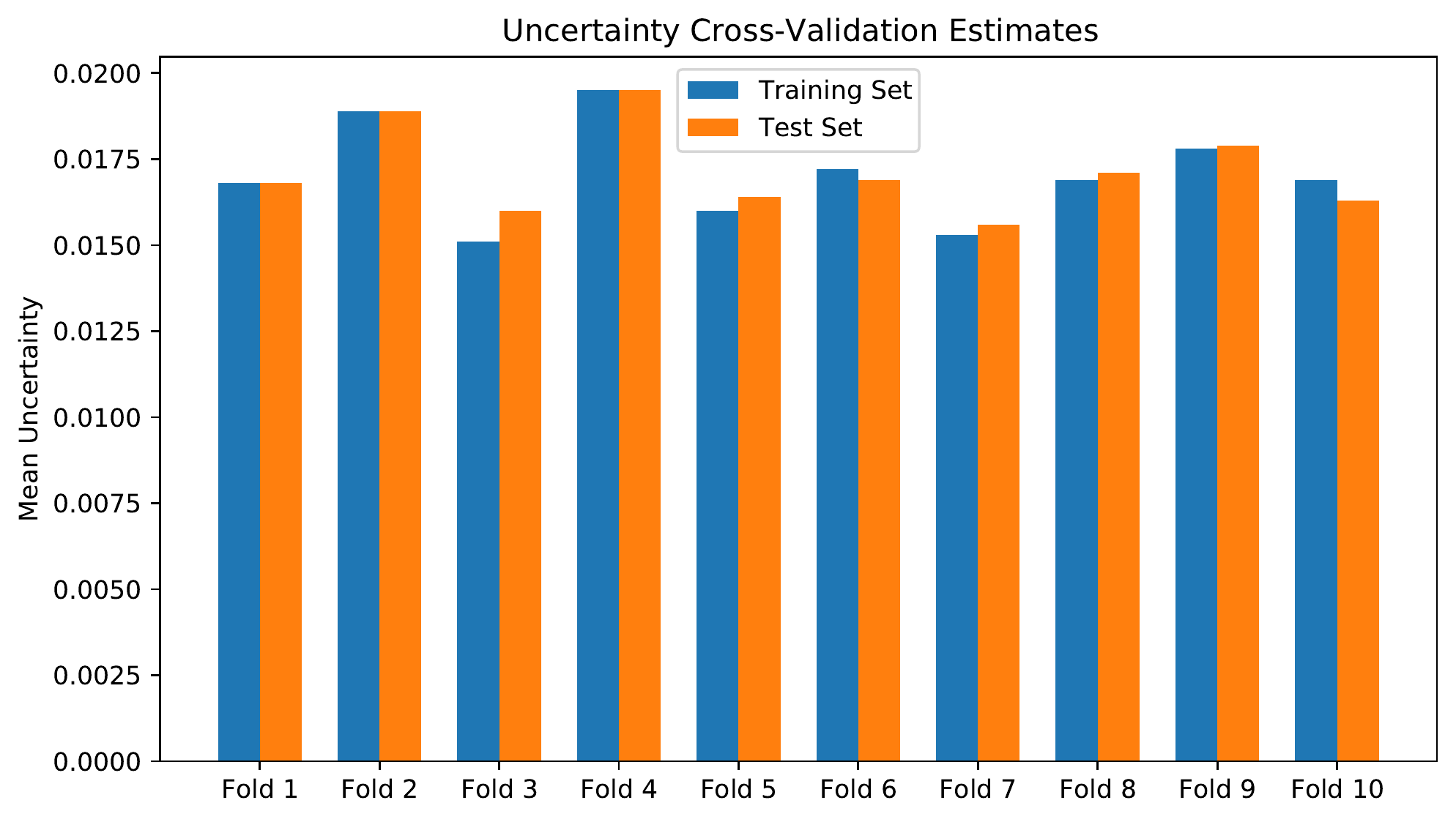}
    \caption{The mean uncertainties for each Cross-Validation Set}
    \label{fig:Uncertainty Estimation Graph}
\end{figure}

The uncertainty was estimated by running each sample 5 times from the dataset through the network to obtain an array of softmax probabilities. Each sample is run 5 times to compromise between effective uncertainty estimation and time to decision. More iterations may yield insignificant changes to the uncertainty at the cost of longer process times. The standard deviation is calculated from these probabilities and reported as the model's measure of uncertainty. The test set included all samples from the cross-validation set, while the training set are randomly selected samples until the same amount of the test set is reached. 

By utilizing the VAE architecture, we can construct a distribution of the output probabilities, and by calculating the standard deviation we can measure of the aleatoric uncertainty of the data. As shown in Fig.~\ref{fig:Uncertainty Estimation Graph}, over half the folds showed higher uncertainty on the newer, similar data present in the test set. The network is able to portray the difference in the learned training data and the new test set data with higher uncertainty. However, the aleatoric uncertainty is much less sensitive to the new data, as the changes from the outputs are much smaller. The similarity of the test data result in smaller differences in the standard deviation compared to the training set.

\section{Conclusion}
In this paper, we proposed a Variational Autoencoder network algorithm that can distinguish and classify an AFib rhythm from a normal rhythm from an ECG signal input. The classification algorithm provides a measure of confidence in the aleatoric uncertainty of the input. Through experimentation and analysis, we showed how the state-of-the-art classification performance can be coupled with uncertainty estimation. Knowing how much confidence can be placed on the model's decision leverages trusting the algorithm with a task, and requiring a physician's re-evaluation, minimizing the mis-classification risk posed by more deterministic networks. 

\bibliographystyle{IEEEtran}
\bibliography{IEEEabrv,root}
\end{document}